\documentclass
[twocolumn,showpacs,preprintnumbers,pre,tightenlines, superscriptaddress,footinbib]{revtex4}%
\usepackage{amssymb}
\usepackage{graphicx}
\usepackage{epsfig}
\usepackage{dcolumn}
\usepackage{bm}
\usepackage{amsmath}
\usepackage{amsfonts}

\begin{document}

\title{Enhancing neural-network performance via assortativity}
\author{Sebastiano de Franciscis\thanks{sebast26@gmail.com}, Samuel Johnson\thanks{samuel@onsager.ugr.es}, and Joaqu\'{\i}n J. Torres\thanks{jtorres@onsager.ugr.es}\\ \small{\textit{Departamento de
      Electromagnetismo y F\'{\i}sica de la Materia, and\\ Institute
      \textit{Carlos I}\ for Theoretical and Computational Physics,\\ Facultad
      de Ciencias, University of Granada, 18071 Granada, Spain.}}}

\begin{abstract}
The performance of attractor neural networks has been shown to depend crucially on the heterogeneity of the underlying topology.
We take this analysis a step further by examining the effect of degree-degree correlations -- or assortativity -- on neural-network behavior.
We make use
of a method recently put forward for studying correlated networks and dynamics thereon, both analytically and computationally, which is independent of how the topology may have evolved. We show how the robustness to noise is greatly enhanced in assortative (positively correlated) neural networks, especially if it is the hub neurons that store the information.
\end{abstract}





\maketitle

\section{Background}
\label{sec_introduction}

For a dozen years or so now, the study of complex systems has
been heavily influenced by
results from network science -- which one might regard as the fusion of graph theory with statistical physics \cite{Newman_rev, Boccaletti}. Phenomena as diverse as epidemics \cite{Watts}, cellular function \cite{Ojalvo}, power-grid failures \cite{Havlin_failures} or internet routing \cite{Boguna_hyperbolic}, among many others \cite{Arenas_rev}, depend crucially on the structure of the underlying network of interactions.
One of the earliest systems to have been described as a network was the brain, which is made up of a great many neurons connected to each other by synapses
\cite{Cajal, Amit, Abbot_from, Torres_rev}. Mathematically, the first neural networks combined the Ising model \cite{Baxter_exact} with the Hebb learning rule \cite{Hebb} to reproduce, very successfully, the storage and retrieval of information \cite{Hopfield, Amari, Amit_Hebb}. Neurons were simplified to binary variables (like Ising spins) representing firing or non-firing cells. By considering the trivial fully-connected topology, exact solutions could be reached, which at the time seemed more important than attempting to introduce biological realism. Subsequent work has tended to focus on considering richer dynamics for the individual cells rather than on the way in which these are interconnected \cite{Vogels, Tores_competition, Torres_rev}. However, the topology of the brain -- whether at the level of neurons and synapses, cortical areas or functional connections -- is obviously far from trivial \cite{Amaral, Sporns_Chialvo, Eguiluz, Arenas_CElegans, Sporns_09, Johnson_JSTAT}.

The number of neighbors a given node in a network has is called its degree, and much attention is paid to degree distributions since they tend to be highly heterogeneous for most real networks. In fact, they are often approximately scale-free (i.e., described by power laws) \cite{Newman_rev, Boccaletti, Peri, Barabasi_cell}. By including this topological feature in a Hopfield-like neural-network model, Torres {\it et al.} \cite{Torres_Influence} found that degree heterogeneity increases the system's performance at high levels of noise, since the hubs (high degree nodes) are able to retain information at levels well above the usual critical noise. To prove this analytically, the authors considered the {\it configurational ensemble} of networks (the set of random networks with a given degree distribution but no degree-degree correlations) and showed that
Monte Carlo (MC) simulations were in good agreement with mean-field analysis, despite the approximation inherent to the latter technique when the network is not fully connected. A similar approach can also be used to show how heterogeneity may be advantageous for the performance of certain tasks in models with a richer dynamics \cite{Johnson_EPL}. It is worth mentioning that this influence of the degree distribution on dynamical behavior is found in many other settings, such as the more general situation of systems of coupled oscillators \cite{Barahona_02}.

Another property of empirical networks that is quite ubiquitous is the existence of correlations between the degrees of nodes and those of their neighbors \cite{Pastor-Satorras, Newman_mixing}. If the average degree-degree correlation is positive the network is said to be {\it assortative}, while it is called {\it disassortative} if negatively correlated. Most heterogeneous networks are disassortative \cite{Newman_rev}, which seems to be because this is in some sense their equilibrium (maximum entropy) state given the constraints imposed by the degree distribution \cite{Johnson_PRL}. However, there are probably often mechanisms at work which drive systems from equilibrium by inducing different correlations, as appears to be the case for most social networks, in which nodes (people) of a kind tend to group together. This feature, known as {\it assortativity} or {\it mixing by degree}, is also relevant for processes taking place on networks. For instance, assortative networks have lower percolation thresholds and are more robust to targeted attack \cite{Newman_mixing}, while disassortative ones make for more stable ecosystems and are -- at least according to the usual definition -- more synchronizable \cite{Brede}.

The approach usually taken when studying correlated networks computationally is to generate a network from the configuration ensemble and then introduce correlations (positive or negative) by some stochastic rewiring process \cite{Maslov}. A drawback of this method, however, is that results may well then depend on the details of this mechanism: there is no guarantee that one is correctly sampling the phase space of networks with given correlations. For analytical work, some kind of hidden variables from which the correlations originate are often considered
\cite{Caldarelli_fitness, Soderberg, Boguna, Fronczak} -- an assumption which can also be used to generate correlated networks computationally \cite{Boguna}. This can be a very powerful method for solving specific network models. However, it may not be appropriate if one wishes to consider all possible networks with given degree-degree correlations, independently of how these may have arisen.
Here we get round this problem by making use of a method recently suggested by Johnson {\it et al.} \cite{Johnson_PRL} whereby the ensemble of all networks with given correlations can be considered theoretically without recurring to hidden variables. Furthermore, we show how this approach can be used computationally to generate random networks that are representative of the ensemble of interest (i.e., they are model-independent). In this way, we study the effect of correlations on a simple neural network model and find that assortativity increases performance in the face of noise -- particularly if
it is the hubs that are mainly responsible for storing information (and it is worth mentioning that there is experimental evidence suggestive of a main functional role played by hub neurons in the brain \cite{Morgan, Bonifazi}).
The
good agreement between the mean-field analysis and our MC simulations bears witness both to the robustness of the results as regards neural systems, and to the viability of using this method for studying dynamics on correlated networks.

\section{Preliminary considerations}

\subsection{Model neurons on networks}
\label{sec_model}

The attractor neural network model put forward by Hopfield \cite{Hopfield} consists of $N$ binary neurons, each with an activity given by the dynamic variable $s_{i}=\pm 1$. 
Every time step (MCS), each neuron is updated according to the stochastic transition probability
$
P(s_{i}\rightarrow \pm 1)=\frac{1}{2}\left[1\pm\tanh\left(h_{i}/T\right)\right]
$
(parallel dynamics),
where the field $h_{i}$ is the combined effect on $i$ of all its neighbors, $h_{i}=\sum_{j}\hat{w}_{ij}s_{j}$, and $T$ is a noise parameter we shall call {\it temperature}, but which represents any kind of random fluctuations in the environment. This is the same as the Ising model for magnetic systems, and the transition rule can be derived from a simple interaction energy such that aligned variables $s$ (spins) contribute less energy than if they were to take opposite values. However, this system can store $P$ given configurations ({\it memory patterns}) $\xi_{i}^{\nu}=\pm 1$ by having the interaction strengths ({\it synaptic weights}) set according to the Hebb rule \cite{Hebb}: $\hat{w}_{ij}\propto \sum_{\nu=1}^{P}\xi_{i}^{\nu}\xi_{j}^{\nu}$. In this way, each pattern becomes an attractor of the dynamics, and the system will evolve towards whichever one is closest to the initial state it is placed in. This mechanism is called {\it associative memory}, and is nowadays used routinely for tasks such as image identification. What is more, it has been established that something similar to the Hebb rule is implemented in nature via the processes of long-term potentiation and depression at the synapses \cite{Ole}, and this phenomenon is indeed required for learning \cite{Gruart}.

To take into account the topology of the network, we shall consider the weights to be of the form $\hat{w}_{ij}=\hat{\omega}_{ij}\hat{a}_{ij}$, where the element $\hat{a}_{ij}$ of the adjacency matrix represents the number of directed edges (usually interpreted as synapses in a neural network) from node $j$ to node $i$, while $\hat{\omega}$ stores the patterns, as before: 
$$
\hat{\omega}_{ij}=\frac{1}{\langle k\rangle}\sum_{\nu=1}^{P}\xi_{i}^{\nu}\xi_{j}^{\nu}.
$$
For the sake of coherence with previous work, we shall assume $\hat{a}$ to be symmetric (i.e., the network is undirected), so each node is characterized by a single degree $k_{i}=\sum_{j}\hat{a}_{ij}$. However, all results are easily extended to directed networks -- in which nodes have both an {\it in} degree, $k_{i}^{\mbox{in}}=\sum_{j}\hat{a}_{ij}$, and an {\it out} degree, $k_{i}^{\mbox{out}}=\sum_{j}\hat{a}_{ji}$ -- by bearing in mind it is only a neuron's pre-synaptic neighbors that influence its behavior. The mean degree of the network is $\langle k\rangle$, where the angles stand for an average over nodes: $\langle \cdot\rangle\equiv N^{-1}\sum_{i}(\cdot)$ \cite{footnote}.


\subsection{Network ensembles}
\label{sec_ensembles}

When one wishes to consider a set of networks which are randomly wired while respecting certain constraints -- that is, an {\it ensemble} -- it is usually useful to define the expected value of the adjacency matrix, $E(\hat{a})\equiv\hat{\epsilon}$ \cite{footnote_macro}. The element $\hat{\epsilon}_{ij}$ of this matrix is the mean value of $\hat{a}_{ij}$ obtained by averaging over the ensemble. For instance, in the Erd\H{o}s-R\'{e}nyi (ER) ensemble all elements (outside the diagonal) take the value $\hat{\epsilon}_{ij}^{ER}=\langle k\rangle /N$, which is the probability that a given pair of nodes be connected by an edge. For studying networks with a given degree sequence, $(k_{1},...k_{N})$, it is common to assume the {\it configuration ensemble}, defined as
$$
\epsilon_{ij}^{conf}=\frac{k_{i}k_{j}}{\langle k\rangle N}
$$
This expression can usually be applied also when the constraint is a given degree distribution, $p(k)$, by integrating over $p(k_{i})$ and $p(k_{j})$ where appropriate. One way of deriving $\hat{\epsilon}^{conf}$ is to assume one has $k_{i}$ dangling half-edges at each node $i$; we then randomly choose pairs of half-edges and join them together until the network is wired up. Each time we do this, the probability that we join $i$ to $j$ is $k_{i}k_{j}/(\langle k\rangle N)^{2}$, and we must perform the operation $\langle k\rangle N$ times. Bianconi showed that this is also the solution for Barab\'asi-Albert evolved networks \cite{Bianconi_mean-field}. However, we should bear in mind that this result is only strictly valid for networks constructed in certain particular ways, such as in these examples. It is often implicitly assumed that were we to average over all random networks with a given degree distribution, the mean adjacency matrix obtained would be $\hat{\epsilon}^{conf}$. As we shall see, however, this is not necessarily the case \cite{Johnson_PRL}.

\begin{figure}[htb!]
\begin{center}
\hspace*{-0.25cm}\includegraphics[width=8.7cm]{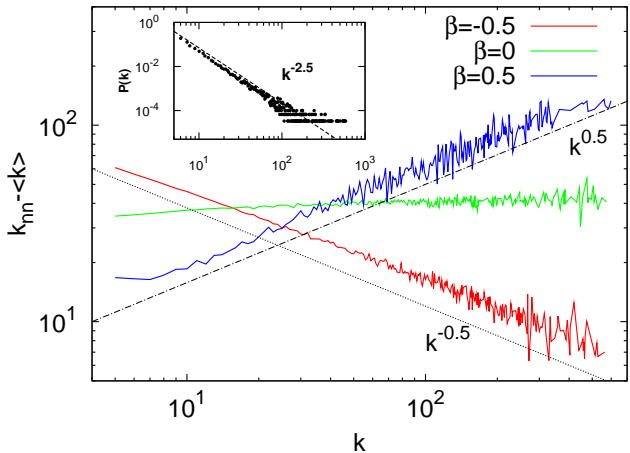}
\end{center}
\caption{Mean-nearest-neighbor functions $\overline{k}_{nn}(k)$ for scale-free networks with $\beta=-0.5$ (disassortative), $0.0$ (neutral), and $0.5$ assortative, generated according to the algorithm described in Sec. \ref{sec_generating}. Inset: degree distribution (the same in all three cases). Other parameters are $\gamma=2.5$, $\langle k\rangle = 12.5$, $N=10^{4}$.}
\label{fig_knn}
\end{figure}


\subsection{Correlated networks}
\label{sec_correlated_nets}

In the configuration ensemble, the expected value
of the mean degree of the neighbors of a given node is $
\overline{k}_{nn,i}=k_{i}^{-1}\sum_{j}\hat{\epsilon}_{ij}^{conf}k_{j}=\langle
k^{2}\rangle/\langle k\rangle,
$ which is independent of $k_{i}$. However, as
mentioned above, real networks often display degree-degree correlations, with the result
that $\overline{k}_{nn,i}=\overline{k}_{nn}(k_{i})$. If $\overline{k}_{nn}(k)$ increases with $k$, the network is said to be assortative -- whereas it is disassortative if it decreases with $k$ (see Fig. \ref{fig_knn}). This is from the more general nomenclature (borrowed form sociology) in which sets are assortative if elements of a kind group together, or assort. In the case of degree-degree correlated networks, positive assortativity means that edges are more than randomly likely to occur between nodes of a similar degree.
A popular measure of this phenomenon is Pearson's coefficient applied to the edges \cite{Newman_rev, Newman_mixing, Boccaletti}: $ r= ([ k_{l}k'_{l}]-[ k_{l}]^{2})/([ k_{l}^{2}]-[ k_{l}]^{2}), $ where $k_{l}$ and $k'_{l}$ are the degrees of each of the two nodes belonging to edge $l$, and $[\cdot]\equiv(\langle k\rangle N)^{-1}\sum_{l}(\cdot)$ is an average over edges.

The ensemble of all networks with a given degree sequence $(k_{1},...k_{N})$
contains a subset for all members of which $\overline{k}_{nn}(k)$ is constant (the configuration ensemble), but also subsets displaying other functions $\overline{k}_{nn}(k)$. We can identify each one of these subsets (regions of phase space) with an
expected adjacency matrix $\hat{\epsilon}$ which simultaneously satisfies the following conditions: ${\bf i)}$  $\sum_{j}k_{j}\hat{\epsilon}_{ij}=k_{i}\overline{k}_{nn}(k_{i})$, $\forall i$ (by definition of $\overline{k}_{nn}(k)$), and ${\bf ii)}$ $\sum_{j}\hat{\epsilon}_{ij}=k_{i}$, $\forall i$ (for consistency). An ansatz which fulfills these requirements is any matrix of the form
\begin{equation}
  \hat{\epsilon}_{ij}=\frac{k_{i}k_{j}}{\langle k\rangle N}
  +\int d\nu \frac{f(\nu)}{N}\left[\frac{(k_{i}k_{j})^{\nu}}
    {\langle k^{\nu}\rangle}-k_{i}^{\nu}-k_{j}^{\nu}+\langle k^{\nu}\rangle  \right],
\label{epsi_gen}
\end{equation}
where $\nu\in\mathbb{R}$ and the function $f(\nu)$ is in general arbitrary \cite{Johnson_PRL}. (If the network were directed, then $k_{i}=k_{i}^{\mbox{in}}$ and $k_{j}=k_{j}^{\mbox{out}}$ in this expression.) 
This ansatz yields
\begin{eqnarray}
  \overline{k}_{nn}(k)=\frac{\langle k^{2}\rangle}{\langle k\rangle}
  +\int d\nu f(\nu)\sigma_{\nu+1}\left[\frac{k^{\nu-1}}
    {\langle k^{\nu}\rangle}-\frac{1}{k} \right]
\label{knn_gen}
\end{eqnarray}
(the first term being the result for the
configuration ensemble), where $\sigma_{b+1}\equiv \langle k^{b+1}\rangle -\langle k\rangle
\langle k^{b}\rangle$.
To prove the uniqueness of a matrix $\hat{\epsilon}$ obtained in this way (i.e., that it is the only one compatible with a given $\overline{k}_{nn}(k)$) assume that there exists another valid matrix $\hat{\epsilon}'\neq\hat{\epsilon}$. Writing $\hat{\epsilon}_{ij}'-\hat{\epsilon}_{ij}\equiv h(k_{i},k_{j})=h_{ij}$, then Condition ${\bf i)}$ implies that $\sum_{j}k_{j}h_{ij}=0$, $\forall i$, while Condition ${\bf ii)}$ means that $\sum_{j}h_{ij}=0$, $\forall i$. It follows that $h_{ij}=0$, $\forall i,j$.
This means that $\hat{\epsilon}$ is not just one possible way of obtaining correlations according to $\overline{k}_{nn}(k)$; rather, there is a two-way mapping between $\hat{\epsilon}$ and $\overline{k}_{nn}(k)$: every network with this particular function $\overline{k}_{nn}(k)$ and no other ones are contained in the ensemble defined by $\hat{\epsilon}$. Thanks to this, if we are able to consider random networks drawn according to this matrix (whether we do this analytically or computationally; see Section \ref{sec_generating}), we can be confident that we are correctly taking account of the whole ensemble of interest. In other words, whatever the reasons behind the existence of degree-degree correlations in a given network, we can study the effects of these with only information on $p(k)$ and $\overline{k}_{nn}(k)$ by obtaining the associated matrix $\hat{\epsilon}$. This is not to say, of course, that all topological properties are captured in this way: a particular network may have other features -- such as higher order correlations, modularity, etc. -- the consideration of which would require concentrating on a sub-partition of those with the same $p(k)$ and $\overline{k}_{nn}(k)$. But this is not our purpose here.
\\
\linebreak

In many empirical networks, $\overline{k}_{nn}(k)$ has the form $\overline{k}_{nn}(k)=A+B
k^{\beta}$, with $A,B>0$ \cite{Boccaletti, Pastor-Satorras} -- the
mixing being assortative if $\beta$ is positive, and disassortative when negative.
Such a case is fitted by Eq. (\ref{knn_gen}) if 
\begin{equation}
f(\nu)=C\left[\frac{\sigma_{2}}{\sigma_{\beta+2}}\delta(\nu-\beta-1)-\delta(\nu-1)\right],
\label{eq_f(nu)}
\end{equation}
with $C$ a positive constant, since this choice yields
\begin{equation}
  \overline{k}_{nn}(k)=\frac{\langle k^{2}\rangle}{\langle k\rangle}
  +C\sigma_{2}\left[\frac{k^{\beta}}{\langle
      k^{\beta+1}\rangle}-\frac{1}{\langle k\rangle} \right].
\label{knn_simple}
\end{equation}


Johnson {\it et al.} \cite{Johnson_PRL} obtained the entropy of ensembles of networks with scale-free degree distributions ($p(k)\sim k^{-\gamma}$) and correlations given by Eq. (\ref{knn_simple}), and found that the most likely configurations (those maximizing the entropy) generally correspond to correlated networks. In particular, the expected mixing, all other things being equal, is usually a certain degree of disassortativity -- which explains the predominance of these networks in the real world. They also showed that the maximum entropy is usually obtained for values of $C$ close to one. Here, we shall use this result to justify concentrating on correlated networks with $C=1$, so that the only parameter we need to take into account is $\beta$. It is worth mentioning that Pastor-Satorras {\it et al.} originally suggested using this exponent as a way of quantifying correlations \cite{Pastor-Satorras}, since this seems to be the most relevant magnitude. Because $\beta$ does not depend directly on $p(k)$ (as $r$ does), and can be defined for networks of any size (whereas $r$, in very heterogeneous networks, always goes to zero for large $N$ due to its normalization \cite{Goltsev_zero}), we shall henceforth use $\beta$ as our assortativity parameter.

So, after plugging Eq. (\ref{eq_f(nu)}) into Eq. (\ref{epsi_gen}), we find that the ensemble of networks exhibiting correlations given by Eq. (\ref{knn_simple}) (and $C=1$) is defined by the mean adjacency matrix
\begin{eqnarray}
& \hat{\epsilon}_{ij} &  =  \frac{1}{N}[k_{i}+k_{j}-\langle k\rangle]
\nonumber
\\
& +& \frac{\sigma_{2}}{\sigma_{\beta+2}}\frac{1}{N}\left[\frac{(k_{i}k_{j})^{\beta+1}}{\langle k^{\beta+1}\rangle} -k_{i}^{\beta+1}-k_{j}^{\beta+1}+\langle k^{\beta+1}\rangle\right].
\label{eq_epsi}
\end{eqnarray}


\section{Analysis and results}

\subsection{Mean field}
\label{sec_asso_dyn}

Let us consider the single-pattern case ($P=1$, $\xi_{i}=\xi_{i}^{1}$). Substituting the adjacency matrix $\hat{a}$ for its expected value $\hat{\epsilon}$ (as given by Eq. (\ref{eq_epsi})) in the expression for the local field at $i$ -- which amounts to a mean-field approximation -- we have
\begin{eqnarray*}
h_{i} & = & \frac{1}{\langle k\rangle}\xi_{i}\left\{ \left[(k_{i}-\langle k\rangle)+\frac{\sigma_{2}}{\sigma_{\beta+2}}(\langle k^{\beta+1}\rangle-k_{i}^{\beta+1})\right]\mu_{0}\right.\\
 & + & \left.\langle k\rangle\mu_{1}+\frac{\sigma_{2}}{\sigma_{\beta+2}}(k_{i}^{\beta}-\langle k^{\beta+1}\rangle)\mu_{\beta+1}\right\},
\end{eqnarray*}
where we have defined
$$
\mu_{\alpha}\equiv \frac{\langle k_{i}^{\alpha} \xi_{i}s_{i}\rangle}{\langle k^{\alpha}\rangle }
$$
for $\alpha=0,$ $1$, $\beta+1$. These order parameters measure the extent to which the system is able to recall information in spite of noise \cite{Johnson_EPL}. For the first order we have $\mu_{0}=m\equiv\langle \xi_{i}s_{i}\rangle$, the standard overlap measure in neural networks (analogous to magnetization in magnetic systems), which takes account of memory performance. However, $\mu_{1}$, for instance, weighs the sum with the degree of each node, with the result that it measures information per synapse instead of per neuron. Although the overlap $m$ is often assumed to represent, in some sense, the {\it mean firing rate} of neurological experiments, it is possible that $\mu_{1}$ is more closely related to the empirical measure, since the total electric potential in an area of tissue is likely to depend on the number of synapses transmitting action potentials. In any case, a comparison between the two order parameters is a good way of assessing to what extent the performance of neurons depends on their degree -- 
larger-degree model neurons can in general store information at higher temperatures than ones with smaller degree can \cite{Torres_Influence}.

Substituting $s_{i}$ for its expected value according to the transition probability, $s_{i}\rightarrow \tanh(h_{i}/T)$, we have, for any $\alpha$,
\[
\langle k_{i}^{\alpha} \xi_{i}s_{i}\rangle=\langle k_{i}^{\alpha} \xi_{i}\tanh(h_{i}/T)\rangle;
\]
or, equivalently, 
the following 3-D map of closed
coupled equations for the macroscopic overlap observables $\mu_{0}$,
$\mu_{1}$ and $\mu_{\beta+1}$ -- which describes, in this mean-field approximation, the dynamics of the system:
\begin{eqnarray}
\mu_{0}(t+1) & = & \int p(k)\tanh[F(t)/(\langle k\rangle T)]\mbox{d}k
\nonumber
\\
\mu_{1}(t+1) & = & \frac{1}{\langle k\rangle}\int p(k)k\tanh[F(t)/(\langle k\rangle T)]\mbox{d}k
\label{eq_3map}
\\
\mu_{\beta+1}(t+1) & = & \frac{1}{\langle k^{\beta+1}\rangle}\int p(k)k^{\beta+1}\tanh[F(t)/(\langle k\rangle T)]\mbox{d}k,
\nonumber
\end{eqnarray}
 with 
\begin{eqnarray*}
F(t) & \equiv & (k\mu_{0}(t)+\langle k\rangle\mu_{1}(t)-\langle k\rangle\mu_{0}(t))\\
& + & \frac{\sigma_{2}}{\sigma_{\beta+2}}[k^{\beta+1}(\mu_{\beta+1}(t)-\mu_{0}(t))\\
& + & \langle k^{\beta+1}\rangle(\mu_{0}(t)-\mu_{\beta+1}(t))].
\end{eqnarray*}
This can be easily computed for any degree distribution $p(k)$. Note that taking $\beta=0$ (the uncorrelated case) the system collapses to the 2-D map obtained in Ref. \cite{Torres_Influence}, while it becomes the typical 1-D case for a homogeneous $p(k)$ -- say a fully-connected network \cite{Hopfield}. It is in principle possible to do similar mean-field analysis for any number $P$ of patterns, but the map would then be $3P$-dimensional, making the problem substantially more complex.

At a critical temperature $T_{c}$, the system will undergo the characteristic second order phase transition from a phase in which it exhibits memory (akin to ferromagnetism) to one in which it does not (paramagnetism). To obtain this critical temperature, we can expand the hyperbolic tangent in Eqs. (\ref{eq_3map}) around the trivial solution $(\mu_{0},\mu_{1},\mu_{\beta+1})\simeq(0,0,0)$ and, keeping only linear terms, write

\begin{eqnarray*}
\mu_{0} & = & \mu_{1}/T_{c}, \\
\mu_{1} & = & \frac{1}{ \langle k\rangle ^{2}T_{c}}\left[ \langle k\rangle ^{2}\mu_{1}+\sigma_{2}\mu_{\beta+1}\right], \\
\mu_{\beta+1} & = & \frac{1}{T_{c} \langle k\rangle  \langle k^{\beta+1}\rangle }\left[\sigma_{\beta+2}\mu_{0}\frac{}{}\right.
\\
& + & \left. \frac{\sigma_{2}}{\sigma_{\beta+2}}\left( \langle k^{\beta+1}\rangle ^{2}- \langle k^{2(\beta+1)}\rangle \right)\mu_{0} \right.\\
            & + & \left.  \langle k\rangle  \langle k^{\beta+1}\rangle \mu_{1}-\frac{\sigma_{2}}{\sigma_{\beta+2}}\left( \langle k^{\beta+1}\rangle ^{2}- \langle k^{2(\beta+1)}\rangle \right)\mu_{\beta+1}\right].
\end{eqnarray*}
Defining
\begin{eqnarray*}
A         & \equiv & \frac{\sigma_{2}}{ \langle k\rangle ^{2}}, \\
B         & \equiv & \frac{\sigma_{2}}{\sigma_{\beta+2}}  \frac{ \langle k^{2(\beta+1)}\rangle - \langle k^{\beta+1}\rangle ^{2}}{ \langle k\rangle  \langle k^{\beta+1}\rangle }, \\
D & \equiv & \frac{\sigma_{\beta+2}}{ \langle k\rangle  \langle k^{\beta+1}\rangle },
\end{eqnarray*}
$T_{c}$ will be the solution to the third order polynomial equation:
\begin{equation}
T_{c}^{3}-(B+1)T_{c}^{2}+(B-A)T_{c}+A(B-D)=0.
\label{eq_PolyTc}
\end{equation}
Note that for neutral (i.e., uncorrelated) networks, $\beta=0$, and so $A=B=D$. We then have $T_{c}=\langle k^{2}\rangle/\langle k\rangle^{2}$, as expected \cite{Johnson_EPL}.

\subsection{Generating correlated networks}
\label{sec_generating}

Given a degree distribution $p(k)$, the ensemble of networks compatible with this constraint and with degree-degree correlations according to Eq. (\ref{knn_simple}) (with some exponent $\beta$) is defined by the mean adjacency matrix $\hat{\epsilon}$ of Eq. (\ref{eq_epsi}) -- as described in Section \ref{sec_correlated_nets} and in Ref. \cite{Johnson_PRL}. Therefore, although there will generally be an enormous number of possible networks in this volume of phase space, we can sample them correctly simply by generating them according to $\hat{\epsilon}$. To do this, first we have to assign to each node a degree drawn from $p(k)$. If the elements of $\hat{\epsilon}$ were probabilities, it would suffice then to connect each pair of nodes $(i,j)$ with probability $\hat{\epsilon}_{ij}$ to generate a valid network. Strictly speaking, $\hat{\epsilon}$ is an expected value, which in certain cases can be greater than one. To get round this, we write a probability matrix $\hat{p}=\hat{\epsilon}/a$ with $a$ some value such that all elements of $\hat{p}$ are smaller than one. If we then take random pairs of nodes $(i,j)$ and, with probability $\hat{p}_{ij}$, place an edge between them, repeating the operation until $\frac{1}{2}\langle k\rangle N$ edges have been placed, the expected value of edges joining $i$ and $j$ will be $\hat{\epsilon}_{ij}$. This method is like the {\it hidden variable} technique
\cite{Boguna} in that edges are placed with a predefined probability (which is why the resulting ensemble is canonical). The difference lies in the fact that in the method here described correlations only depend on the degrees of nodes.

We are interested here in neural networks, in which a given pair of nodes can be joined by several synapses, so we shall not impose the restriction of so-called simple networks of allowing only one edge at most per pair. We shall, however, consider networks with a {\it structural cutoff}: $k_{i}< \sqrt{\langle k\rangle N}$, $\forall i$ \cite{Bianconi}. This ensures that, at least for $\beta\leq 0$, all elements of $\hat{\epsilon}$ are indeed smaller than one.



Because we can expect effects due to degree-degree correlations to be largest when $p(k)$ is very broad, and since most networks in nature and technology seem to exhibit approximately power-law degree distributions \cite{Newman_rev, Arenas_rev, Peri, Barabasi_cell}, we shall here test our general theoretical results against simulations of scale-free networks: $p(k)\sim k^{-\gamma}$.
This means that a network (or the region of phase space to which it belongs) is characterized by the set of parameters $\lbrace \langle k\rangle, N, \gamma, \beta \rbrace$.

\subsection{Assortativity and dynamics}
\label{sec_results}

\begin{figure}[htb!]
\begin{center}
\hspace*{-0.2cm}\includegraphics[width=8.7cm]{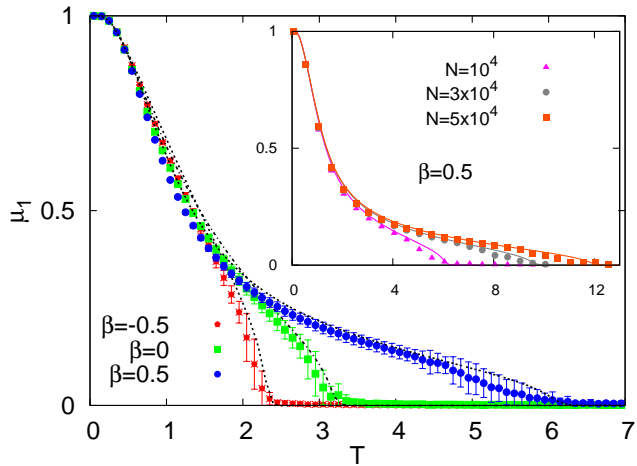}
\end{center}
\caption{Stable stationary value of the weighted overlap $\mu_{1}$ against temperature $T$ for scale-free networks
with correlations according to $\overline{k}_{nn}\sim k^{\beta}$, for $\beta=-0.5$ (disassortative), $0.0$ (neutral), and $0.5$ (assortative). Symbols from MC simulations, with errorbars representing standard deviations, and lines from
Eqs. (\ref{eq_3map}). Other network  parameters as in Fig. \ref{fig_knn}. Inset: $\mu_{1}$ against $T$ for the assortative case ($\beta=0.5$) and different system sizes: $N=10^{4}$, $3\cdot 10^{4}$ and $5\cdot 10^{4}$.}
\label{fig_mu1}
\end{figure}

\begin{figure}[htb!]
\begin{center}
\hspace*{-0.65cm}\includegraphics[width=9.45cm]{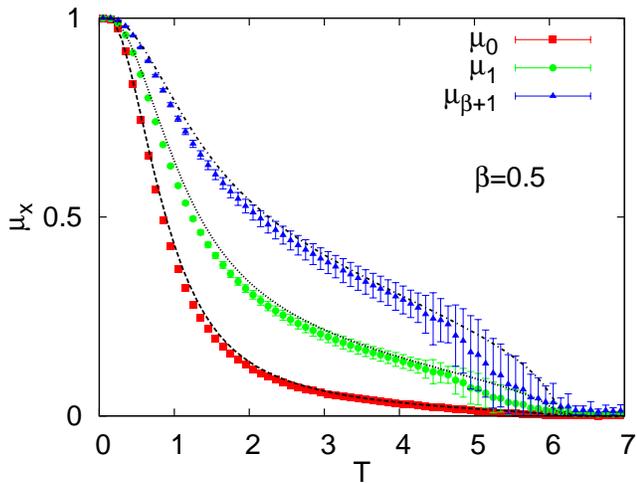}
\end{center}
\caption{Stable stationary values of order parameters $\mu_{0}$, $\mu_{1}$ and $\mu_{\beta+1}$ against temperature $T$, for assortative networks according to $\beta=0.5$.
Symbols from MC simulations, with errorbars representing standard deviations, and lines from Eqs. (\ref{eq_3map}). Other parameters as in Fig. \ref{fig_knn}.
}
\label{fig_mub}
\end{figure}

\begin{figure}[htb!]
\begin{center}
\hspace*{-0.65cm}\includegraphics[width=9.3cm]{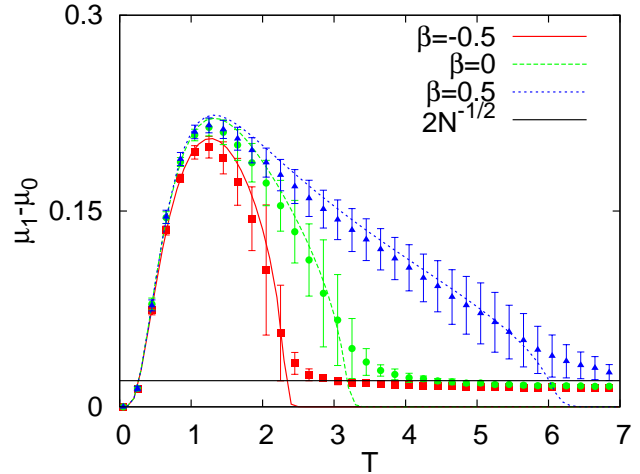}
\end{center}
\caption{
Difference between the stationary values $\mu_{1}$ and $\mu_{0}$ for networks with $\beta=-0.5$ (disassortative), $0.0$ (neutral) and $0.5$ (assortative), against temperature.
Symbols from MC simulations, with errorbars representing standard deviations, and lines from Eqs. (\ref{eq_3map}). Line shows the expected level of fluctuations due to noise, $\sim N^{-\frac{1}{2}}$. Other parameters as in Fig. \ref{fig_knn}.
}
\label{fig_picos}
\end{figure}

In Fig. \ref{fig_mu1} we plot the stationary value of $\mu_{1}$ against the temperature $T$, as obtained from simulations and Eqs. (\ref{eq_3map}),
for disassortative, neutral and assortative networks. The three curves are similar at low temperatures, but as $T$ increases their behavior becomes quite different. The disassortative network is the least robust to noise. However, the assortative one is capable of retaining some information at temperatures considerably higher than the critical value, $T_{c}=\langle k^{2}\rangle /\langle k\rangle$, of neutral networks. A comparison between $\mu_{1}$ and $\mu_{0}$ (see Fig. \ref{fig_mub}) shows that it is the high degree nodes that are mainly responsible for this difference in performance. This can be seen more clearly in Fig. \ref{fig_picos}, which displays the difference $\mu_{1}-\mu_{0}$ against $T$ for the same networks. It seems that, because in an assortative network a sub-graph of hubs will have more edges than in a disassortative one, it has a higher effective critical temperature. Therefore, even when most of the nodes are acting randomly, the set of nodes of sufficiently high degree nevertheless displays associative memory.


\begin{figure}[htb!]
\begin{center}
\hspace*{-0.5cm}\includegraphics[width=9.3cm]{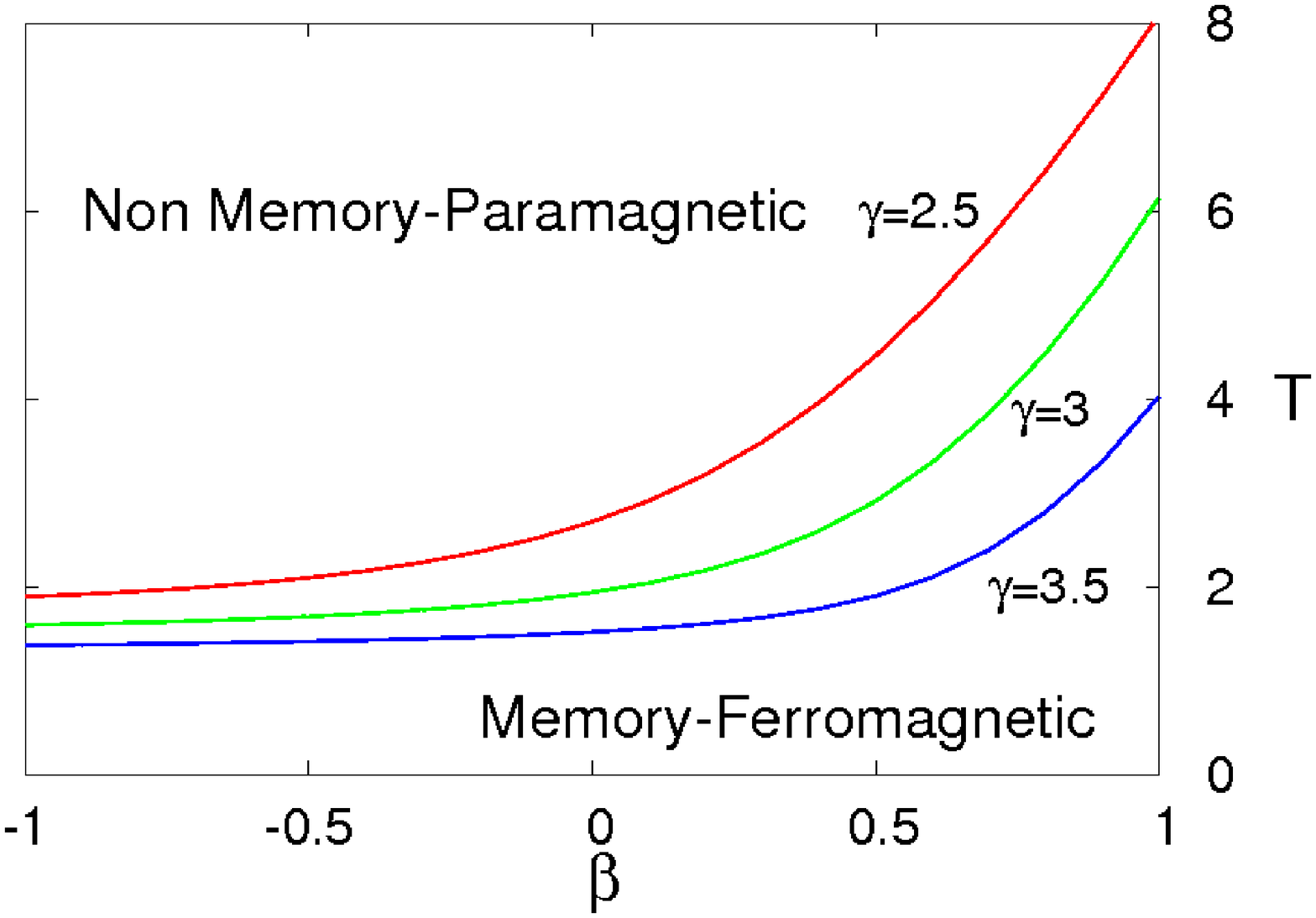}
\end{center}
\caption{Phase diagrams for scale-free networks with $\gamma=2.5$, $3$, and $3.5$. Lines show the critical temperature $T_{c}$ marking the second-order transition from a memory (ferromagnetic) phase to a memoryless (paramagnetic) one, against the assortativity $\beta$, as
given by Eq. (\ref{eq_PolyTc}). Other parameters as in Fig. \ref{fig_knn}.
}
\label{fig_phase}
\end{figure}

The phase diagram if Fig. \ref{fig_phase} shows the critical temperature, $T_{c}$, as obtained from Eq. (\ref{eq_PolyTc}). In addition to the effect reported in Ref. \cite{Torres_Influence} whereby the $T_{c}$ of scale-free networks grows with degree heterogeneity (decreasing $\gamma$), it also increases very significantly with positive degree-degree correlations (increasing $\beta$).

\begin{figure}[h!]
\begin{center}
\hspace*{-0.5cm}\includegraphics[width=9.4cm]{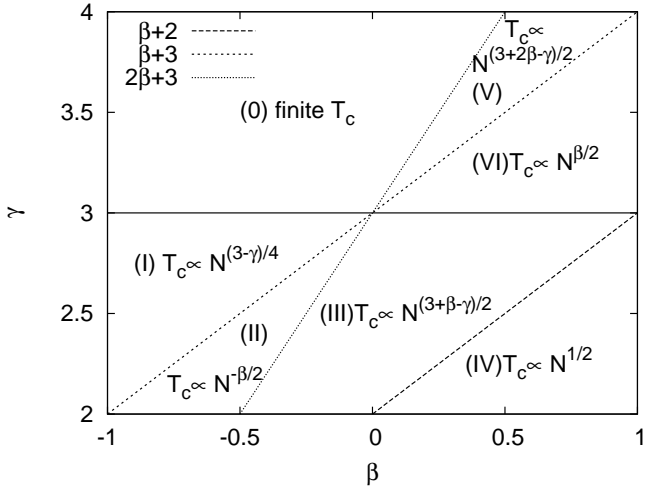}
\caption{Parameter space $\beta-\gamma$ partitioned into the regions in which $b(\beta,\gamma)$ has the same functional form -- where $b$ is the scaling exponent of the critical temperature: $T_{c}\sim N^b$. Exponents obtained by taking the large $N$ limit in Eq. (\ref{eq_PolyTc}).}
\label{fig_scale}
\end{center}
\end{figure}

\begin{figure}[h!]
\begin{center}
\hspace*{-0.3cm}\includegraphics[width=9.1cm]{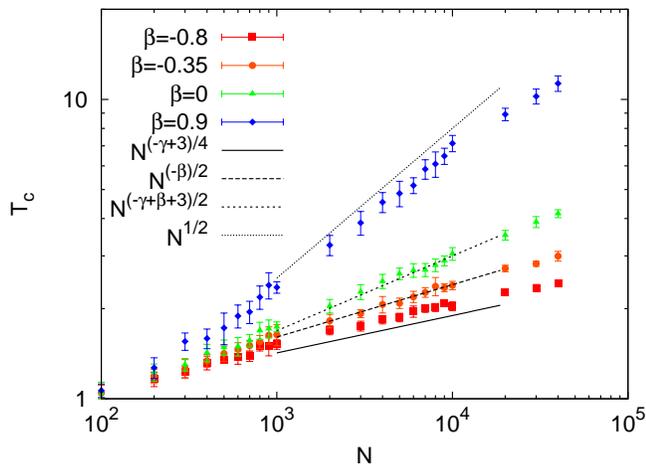}
\caption{
Examples of how $T_{c}$ scales with $N$ for networks belonging to regions I, II, III and IV of Fig. \ref{fig_scale} ($\beta=-0.8$, $-0.35$, $0.0$ and $0.9$, respectively). Symbols from MC simulations, with errorbars representing standard deviations, and slopes from Eq. (\ref{eq_PolyTc}). All parameters -- except for $\beta$ and $N$ -- are as in Fig. \ref{fig_knn}.
}
\label{fig_4lines}
\end{center}
\end{figure}

At large values of $N$, the critical temperature scales as $T_{c}\sim N^b$, with $b\geq0$ a constant. However, because the moments of $k$ appearing in the coefficients of Eq. (\ref{eq_PolyTc}) can have different asymptotic behavior depending on the values of $\gamma$ and $\beta$, the scaling exponent $b$ differs from one region to another in the space of these parameters. These are the seven regions shown in Fig. \ref{fig_scale}, along with the scaling behavior exhibited by each one. This can be seen explicitely in Fig. \ref{fig_4lines}, where $T_{c}$, as obtained from MC simulations, is plotted against $N$ for cases in each of the regions with $\gamma<3$. In each case, the scaling is as given by Eq. (\ref{eq_PolyTc}) and shown in Fig. \ref{fig_scale}.
For the four regions with $\gamma<3$, from lowest to highest assortativity we have scaling exponents which are dependent on: only $\gamma$ (region I), only $\beta$ (region II), both $\gamma$ and $\beta$ (region III), and, perhaps most interestingly, neither of the two (region IV) -- with $T_{c}$ scaling, in the latter case, as $\sqrt{N}$. As for the more homogeneous $\gamma>3$ part, regions V and VI have a diverging critical temperature despite the fact that the second moment of $p(k)$ is finite, simply as a result of assortativity.

\section{Discussion}
\label{sec_discussion}

We have shown that assortative networks of simple model neurons are able to exhibit associative memory in the presence of levels of noise such that uncorrelated (or disassortative) networks cannot. This may appear to be in contradiction with a recent result obtained using spectral graph analysis -- that synchronizability of a set of coupled oscillators is highest for disassortative networks \cite{Brede}. A synchronous state of model oscillators and a memory phase of model neurons are both sets of many simple dynamical elements coupled via a network in such a way that a macroscopically coherent situation is maintained \cite{Barahona_02}. Obviously both systems require the effective transmission of infomation among the elements. So why are opposite results as regards the influence of topology reported for each system? The answer is simple: whereas the definition of a synchronous state is that every single element oscillate at the same frequency, it is precisely when most elements are actually behaving randomly that the advantages to assortativity we report become apparent. In fact, it can be seen in Fig. \ref{fig_mu1} that at low temperatures disassortative networks perform the best, although the effect is small. This is reminiscent of percolation: at high densities of edges the giant component is larger in disassortative networks, but in assortative ones a non-vanishing fraction of nodes remain interconnected even at densities below the usual percolation threshold \cite{Newman_mixing}. Because in the case of targeted attacks it is this threshold which is taken as a measure of resilience, we say that assortative networks perform the best. In general,
the optimal network for good conditions (i.e., complete synchronization, high density of edges, low levels of noise) is not necessarily the one which performs the best in bad conditions (partial synchronization, low density of edges, high levels of noise). It seems that optimality --  whether in resilience or robustness -- should thus be defined for particular conditions.

We have used the technique suggested in Ref. \cite{Johnson_PRL} to study the effect of correlations on networks of model neurons, but many other systems of dynamical elements should be susceptible to a similar treatment. In fact, Ising spins \cite{Bianconi_mean-field}, Voter Model agents \cite{Eguiluz_voter}, or Boolean nodes \cite{Tiago}, for instance, are similar enough to binary neurons that we should expect similar results for these models. If a moral can be drawn, it is that persistence of partial synchrony, or coherence of a subset of highly connected dynamical elements, can sometimes be as relevant (or more so) as the possibility of every element behaving in the same way. In the case of real brain cells,
experiments suggest that hub neurons play key functional roles \cite{Morgan, Bonifazi}.
From this point of view, there may be a selective pressure for brain networks to become assortative -- although, admittedly, this organ engages in such complex behavior that there must be many more functional constraints on its structure than just a high robustness to noise. Nevertheless, it would be interesting to investigate this aspect of biological systems experimentally. For this, it should be borne in mind that heterogeneous networks have a natural tendency to become disassortative, so it is against the expected value of correlations discussed in Ref. \cite{Johnson_PRL} that empirical data should be contrasted in order to look for meaningful deviations towards assortativity.
\\
\linebreak

\acknowledgments
{
This work was supported by Junta de Andaluc\'{i}a projects FQM-01505 and P09-FQM4682,
and by Spanish MEC-FEDER project FIS2009-08451. Many thanks to Omar Al Hammal for fruitful suggestions, and to Antani Tarapiatapioca for inspiration as well as practical advice.
}

\end{document}